\documentclass[]{spie}  %>>> use for US letter paper
\usepackage{amsmath,amsfonts,amssymb}
\usepackage{graphicx}
\usepackage{booktabs}
\usepackage[colorlinks=true, allcolors=blue]{hyperref}
\usepackage{cleveref}
\usepackage[subrefformat=parens,farskip=0pt,justification=centering]{subfig}
\usepackage{tikz}
\newcommand*\circled[1]{\tikz[baseline=(char.base)]{\node[shape=circle,draw,inner sep=2pt] (char) {#1};}}

\graphicspath{{./figs/}}

\usepackage{bbding}
\usepackage{multirow}
\usepackage{expl3}
\makeatletter
\ams@newcommand{\multiint}[1]{\DOTSI\protect\MultiIntegral{#1}}
\renewcommand{\MultiIntegral}[1]{%
  \edef\ints@c{\noexpand\intop
    \ifnum#1=\z@\noexpand\intdots@\else\noexpand\intkern@\fi
    \replicate{#1-2}{\noexpand\intop\noexpand\intkern@}%
    \noexpand\intop
    \noexpand\ilimits@
  }%
  \futurelet\@let@token\ints@a
}
\makeatother
\ExplSyntaxOn
\cs_new:Npn \replicate #1 #2 { \prg_replicate:nn { #1 } { #2 } }
\ExplSyntaxOff
\title{Open-Source Differentiable Lithography Imaging Framework}

\author[1,2]{Guojin Chen}
\author[3]{Hao Geng}
\author[1]{Bei Yu}
\author[2]{David Z. Pan}

\affil[1]{The Chinese University of Hong Kong}
\affil[2]{The University of Texas at Austin}
\affil[3]{ShanghaiTech University}

\begin{document}
\maketitle

\begin{abstract}
    The rapid evolution of the electronics industry, driven by Moore's law and the proliferation of integrated circuits, has led to significant advancements in modern society, including the Internet, wireless communication, and artificial intelligence (AI).
    Central to this progress is optical lithography, a critical technology in semiconductor manufacturing that accounts for approximately 30\% to 40\% of production costs.
    As semiconductor nodes shrink and transistor numbers increase, optical lithography becomes increasingly vital in current integrated circuit (IC) fabrication technology.
    This paper introduces an open-source differentiable lithography imaging framework that leverages the principles of differentiable programming and the computational power of GPUs to enhance the precision of lithography modeling and simplify the optimization of resolution enhancement techniques (RETs).
    The framework models the core components of lithography as differentiable segments, allowing for the implementation of standard scalar imaging models, including the Abbe and Hopkins models, as well as their approximation models.
    The paper introduces a computational lithography framework that optimizes semiconductor manufacturing processes using advanced computational techniques and differentiable programming.
    It compares imaging models and provides tools for enhancing resolution, demonstrating improved semiconductor patterning performance.
    The open-sourced framework represents a significant advancement in lithography technology, facilitating collaboration in the field.
    The source code is available at \url{https://github.com/TorchOPC/TorchLitho}.
\end{abstract}

% Include a list of keywords after the abstract

\section{INTRODUCTION}
\label{sec:intro}  % \label{} allows reference to this section

\begin{figure}[tb!]
  \centering
  \subfloat[]{\includegraphics[width=.26\linewidth]{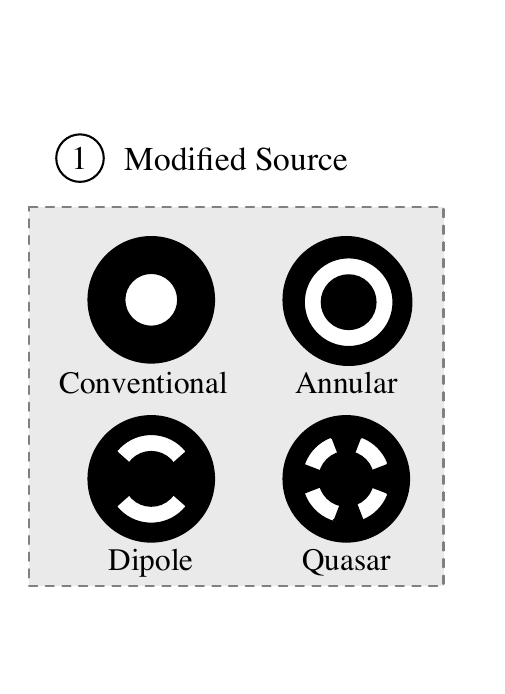} \label{fig:param_source}} \hspace{.02\linewidth}
  \subfloat[]{\includegraphics[width=.38\linewidth]{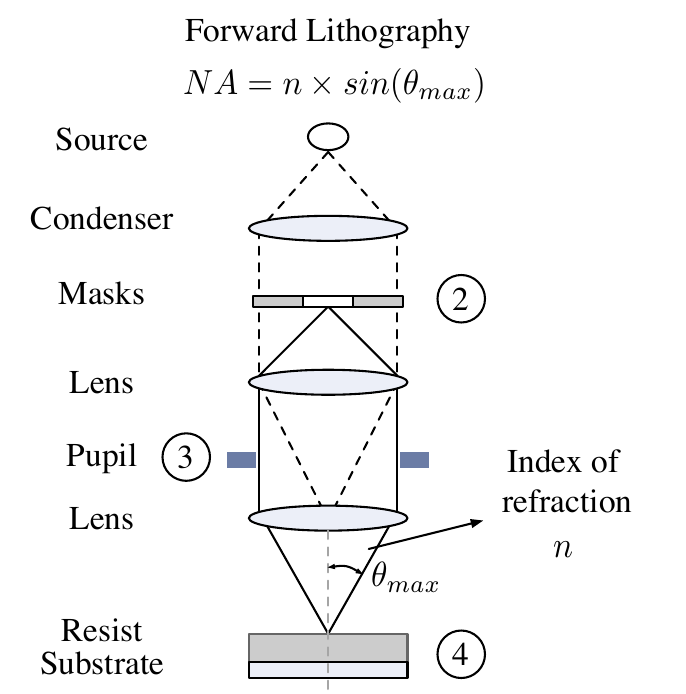}     \label{fig:litho_flow}}   \hspace{.02\linewidth}
  % \subfloat[]{\includegraphics[width=.20\linewidth]{litho/dchain} \label{fig:dchain}}       \hspace{.02\linewidth}
  \caption{
    (a) Different representations of illuminator source, conventional source, annular source, dipole source, ans quasar source.
    (b) Core components of forward lithography process.
    % (c) The visualization of the differentiable lithography chain.
  }
  \label{fig:forward_litho}
\end{figure}

The evolution of the electronics industry, catalyzed by Moore's law and the proliferation of integrated circuits,
has engendered profound paradigm shifts within modern society.
Innovations such as the advent of the Internet, the ubiquity of wireless communication,
and state-of-the-art (SOTA) advancements in artificial intelligence (AI) can be attributed to the formidable computational capacities inherent in IC processors.
At the core of the semiconductor manufacturing industry, optical lithography holds a substantial share of around 30\% to 40\% in production costs~\cite{ma2011computational}.
The progression and efficacy of lithography significantly shape the reduction of critical dimensions in integrated circuits,
influencing transistor speed and silicon real estate.
As advanced semiconductor nodes shrink feature sizes, transistor numbers surge,
thrusting optical lithography into a paramount role within current IC fabrication technology.
As illustrated in \Cref{fig:forward_litho}, a common optical lithography setup comprises four primary elements:
\circled{1} An adjustable illumination system, as presented in \Cref{fig:param_source}.
\circled{2} The photomask setup with a basic binary design made of see-through and non-see-through sections.
\circled{3} An exposure mechanism.
\circled{4} A resist station to yield the end product – the wafer, as illustrated in \Cref{fig:litho_flow}.
The numerical aperture (NA) is a fundamental non-dimensional metric that indicates the span of angles from which the optical setup can receive or radiate light.
It determines the light-absorbing capacity and the clarity of optical setups. When light strikes a photomask, it experiences a process termed as diffraction.
These diffracted beams are then gathered by precision-designed projection lenses and directed towards the surface layered with photoresist.
The resulting chemical reactions in the photoresist, primarily after heating, make certain areas of the resist dissolve in particular developer solutions, marking the essential chemical phase of the lithographic process.
Computational lithography utilizes comprehensive mathematical theories, including inverse problems, mathematical refinement, and computational imaging, to forge optimization-driven resolution improvement methods for optical lithography.
The discipline of lithography, deeply anchored in optical and chemical principles, can be thoroughly defined through complex mathematical models.
In today's lithography landscape, computational approaches are crucial.
With the advancement of GPU and machine learning technologies, a series of works have emerged that utilize GPU~\cite{DATE21_GPU_LevelSet,OPC-ICCAD2021-DevelSet} or machine learning (ML)~\cite{DAC23_Nitho,ICCAD22_AdaOPC,L2OILT_TCAD,liu2022adversarial,large_scale_phys_litho23,TCAD_DAMO,Zhu22DNF} to accelerate lithography.
They entail the use of advanced computational resources to imitate, and more critically, enhance both the optical and chemical mechanisms inherent to lithography.
The aim goes beyond merely mirroring actual processes, aiming to promote growth in lithographic simulation and define the best processing parameters.

\begin{figure}[tb!]
  \centering
  \includegraphics[width=.6\linewidth]{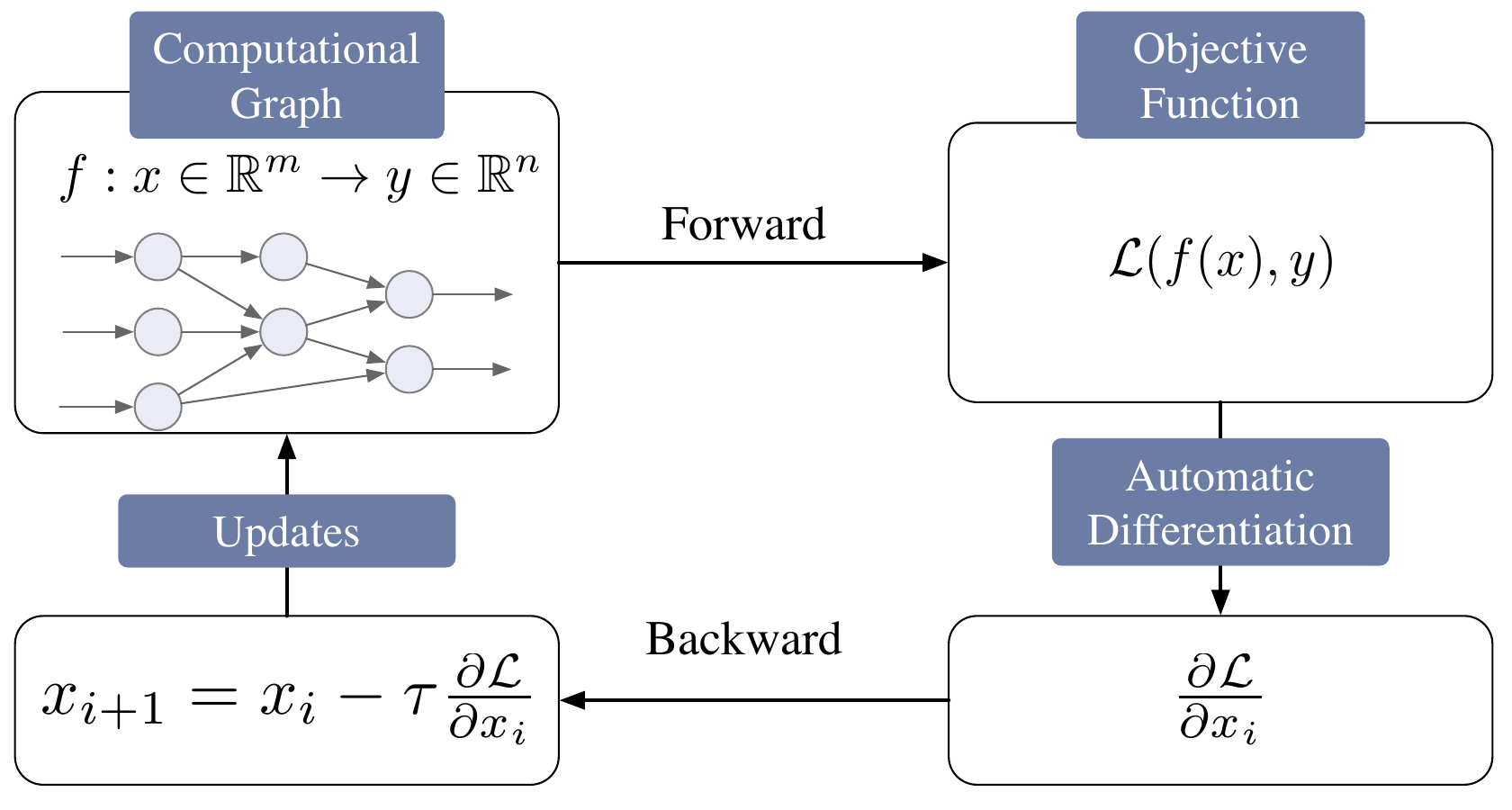}
  \caption{Automatic differentiation flow.}
  \label{fig:diff_programming}
\end{figure}

\textbf{Differentiable programming}, introduced as ``Differentiable Functional Programming'' in 2015~\cite{olah2015neural}, has been gaining traction in recent times.
It's quickly establishing itself as a significant domain and is frequently considered the future of software design. It takes forward the ideas of deep learning, inspired by the rising popularity of machine learning frameworks like TensorFlow~\cite{tensorflow}, PyTorch~\cite{pytorch}, and JAX~\cite{jax2018github}. In essence, differentiable programming is a programming style that involves software made of \textit{differentiable and adjustable} units (or a computation graph). These units undergo automatic differentiation and are fine-tuned to execute a particular function~\cite{chen2023differentiable}. These programs can even adjust sections of their own code based on gradient information.
This coding philosophy has proven its worth in various research areas,~\cite{dreamplace,dreamplace3,ICCAD23:AlphaSyn,ECCV22_hilbert,ISPD21_Pointcloud} especially in imaging,
showcasing its potential in resolving challenges faced by current computational lithography techniques~\cite{ICCAD20_DAMO}.
As depicted in \Cref{fig:diff_programming}, two modes exist for performing automatic differentiation: forward and backward (or reverse).
The differentiation between these methods lies in the sequence in which the derivatives are determined using the chain rule.
The forward method computes numerical derivatives concurrently with function evaluation, progressing from input to result.
In contrast, the backward method expands on the forward computation graph, determining the gradient by navigating the graph in the opposite direction, from result to input.

\section{Algorithm}
\label{sec:algo}  % \label{} allows reference to this section

\textbf{Lithography imaging models}.
The scalar imaging equation under partially coherent illumination can be stated as follows,
\begin{equation}
  \begin{aligned}
  {I}(x, y)= \multiint{6}_{-\infty}^{\infty}  &{J}(f, g) O(f^{\prime}, g^{\prime})  O^{*}(f^{\prime \prime}, g^{\prime \prime})
  {H}(f+f^{\prime}, g+g^{\prime}) {H}^*(f+f^{\prime \prime}, g+g^{\prime \prime}) \\
  & \exp(-j 2 \pi((f^{\prime}-f^{\prime \prime}) x+ (g^{\prime}-g^{\prime \prime}) y))
  \mathrm{~d}f \mathrm{~d} g \mathrm{~d} f^{\prime} \mathrm{~d} g^{\prime} \mathrm{~d} f^{\prime \prime} \mathrm{~d} g^{\prime \prime},
  \end{aligned}
  \label{eq:litho_imaging_full}
\end{equation}
where $O$ is the object plane, the field of the photomask $M$ in the lithography case.
$H$ is the projector transfer function and $J$ is the mutual intensity given by the propagation of all contributing source points of the illuminator.
$I$ is the final aerial image intensity and $(x, y)$, $(f, g)$ are coordinates on spatial and frequency domain.
Assuming the Kohler illumination as our base, we've put into practice two popular models of partially coherent imaging, as referenced in~\cite{fuhner2014artificial}. The initial model is based on Abbe's framework, while the subsequent one adopts the Hopkins diffraction model. It's noteworthy that the Hopkins diffraction model is essentially a streamlined and estimated adaptation of the Abbe's model.
Concurrently, we've also introduced a comprehensive neural network-driven model for lithography imaging.
Abbe's model can be mathematically expressed as the integration over the source.
\begin{equation}
  \begin{aligned}
  I\left(x_1, y_1\right)=\iint_{-\infty}^{\infty} J_C(f, g)\left(\iiiint_{-\infty}^{\infty}\right. & O\left(f^{\prime}, g^{\prime}\right) O^*\left(f^{\prime \prime}, g^{\prime \prime}\right)
  H\left(f+f^{\prime}, g+g^{\prime}\right) H^*\left(f+f^{\prime \prime}, g+g^{\prime \prime}\right) \\
  & \left.\exp \left(-i 2 \pi\left(\left(f^{\prime}-f^{\prime \prime}\right) x_1+\left(g^{\prime}-g^{\prime \prime}\right) y_1\right)\right) \mathrm{d} f^{\prime} \mathrm{d} g^{\prime} \mathrm{d} f^{\prime \prime} \mathrm{d} g^{\prime \prime}\right) \mathrm{d} f \mathrm{~d} g .
  \end{aligned}
\end{equation}
By setting
\begin{equation}
  A_{(f, g)}^{\prime}\left(f^{\prime}, g^{\prime}\right):= H\left(f+f^{\prime}, g+g^{\prime}\right) O\left(f^{\prime}, g^{\prime}\right),
\end{equation}
and perform inverse Fourier transformation, we obtain Abbe's formulation,
\begin{equation}
  \begin{aligned}
  I\left(x_1, y_1\right) =\iint_{-\infty}^{\infty} J_C(f, g)\left|A_{(f, g)}^{\prime}\left(x_1, y_1\right)\right|^2 \mathrm{~d} f \mathrm{~d} g .
  \end{aligned}
  \label{eq:abbe_short}
\end{equation}

Hopkins' diffraction model is a simplified and approximate version of the Abbe's model,
where the integration over the source is carried out before summing up the diffraction angles accepted by the lens.
\begin{equation}
  \begin{aligned}
    {I}(x_1, y_1) =& \iiiint_{-\infty}^{\infty}\mathcal{T}(f^{\prime},g^{\prime};f^{\prime \prime},g^{\prime \prime}) O(f^{\prime},g^{\prime}) O^*(f^{\prime \prime},g^{\prime \prime}) \\
    &\exp (-j 2 \pi((f^{\prime}-f^{\prime \prime}) x_1+(g^{\prime}-g^{\prime \prime}) y_1)) \mathrm{d} f^{\prime} \mathrm{d} g^{\prime} \mathrm{d} f^{\prime \prime} \mathrm{d} g^{\prime \prime},
  \end{aligned}
  \label{eq:aerial_imaging_tcc}
\end{equation}
where $\mathcal{T}$ is the \textit{transmission cross-coefficients} ($TCC$) given by:
\begin{equation}
    {TCC}(f^{\prime}, g^{\prime};f^{\prime \prime}, g^{\prime \prime})=\iint_{-\infty}^{\infty} {J}(f, g)
    {H}(f+f^{\prime}, g+g^{\prime}) {H}^*(f+f^{\prime \prime},g+g^{\prime \prime}) \mathrm{d} f \mathrm{d} g .
\end{equation}
To simplify the computational complexity of the Hopkins imaging equations, a method known as the \textit{Sum of Coherent Source} (SOCS) has been developed. This technique involves breaking down the ${TCC}$ spectrum using Singular Value Decomposition (SVD). The process is mathematically represented as:
\begin{equation}
  {TCC}\left(f^{\prime}, g^{\prime} ; f^{\prime \prime}, g^{\prime \prime}\right)=\sum_{q=1}^{\infty} \kappa_q \Phi_q\left(f^{\prime}, g^{\prime}\right) \Phi_q^*\left(f^{\prime \prime}, g^{\prime \prime}\right).
  \label{eq:tcc_deco}
\end{equation}
In this equation, $\kappa_q$ and ${\Phi_q}$ represent the $q$'th eigenvalue and eigenvector of the ${TCC}$, respectively. Given the rapid decrease of the eigenvalue $\kappa_q$ with increasing $q$, it's feasible to retain only the $Q$ largest eigenvalues to expedite the calculation process. By integrating \Cref{eq:tcc_deco} into \Cref{eq:aerial_imaging_tcc} and applying the Inverse Fast Fourier Transform (\texttt{IFFT}), the SOCS can be reformulated for spatial positioning as shown in \Cref{eq:socs_spatial}:
\begin{equation}
  {I}(x, y)=\sum_{q=1}^Q \kappa_q\left|{\phi}_q(x, y) \otimes {M}(x, y)\right|^2.
  \label{eq:socs_spatial}
\end{equation}
Here, ${\phi}_q(x, y)$ and ${M}(x, y)$ denote the spatial distributions of ${\Phi_q}$ and $M$, respectively.

\textbf{Abbe vs. Hopkins}
\begin{figure}
  \centering
  \includegraphics[width=.99\linewidth]{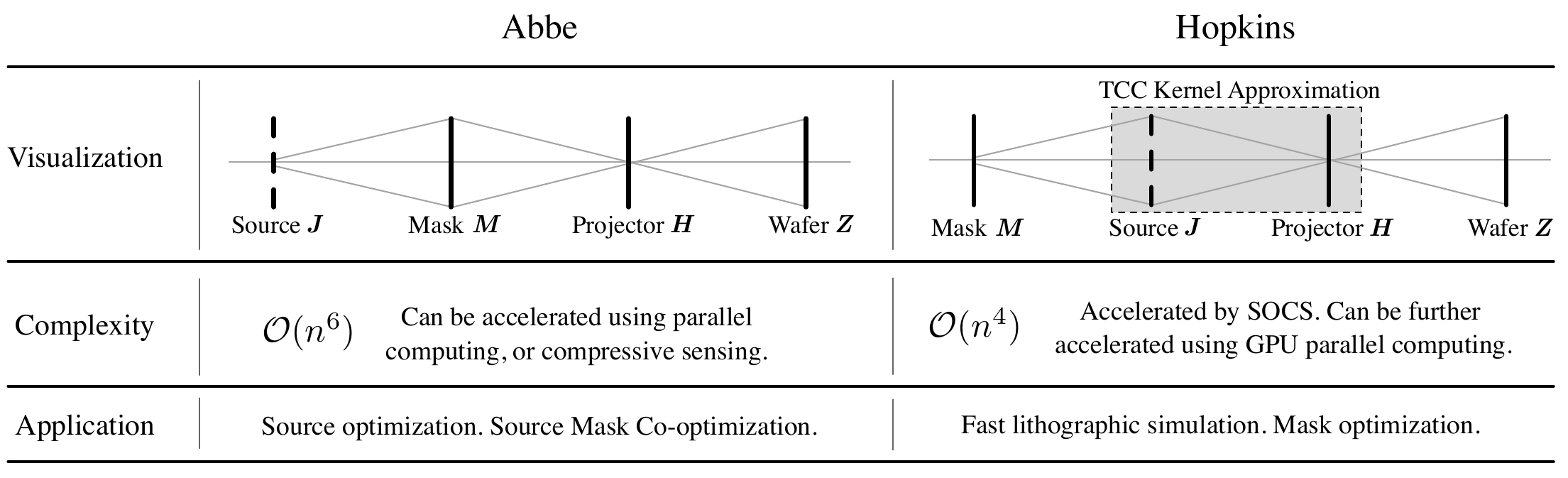}
  \caption{Comparison of Abbe's and Hopkins' imaging.}
  \label{fig:abbe-vs-hopkins}
\end{figure}
As depicted in \Cref{fig:abbe-vs-hopkins}, in the approach outlined by Hopkins (\Cref{eq:socs_spatial}), the SOCS is employed to reduce the computational load from a scale of $\mathcal{O}(n^6)$ down to $\mathcal{O}(Q\times n^4)$.
This technique is notable for its separation of the mask from the optical system, enhancing its suitability for image simulation in conjunction with mask optimization.
Under fixed optical imaging conditions, Hopkins' method demonstrates superior efficiency over Abbe's approach in terms of processing time, which has led to its preference
in various MO algorithms as cited in multiple studies~\cite{OPC-DAC2014-Gao,OPC-DAC2018-Yang,OPC-ICCAD2021-DevelSet,a2ilt2022wang,multiilt2023sun}.
On the other hand, Abbe's method, as detailed in \Cref{eq:abbe_short},
accumulates the effects of all source points to produce the final aerial image.
This approach is naturally better suited for optimizing the source owing to the discretization of the source. Therefore, for effective SO and SMO, the foundation of a lithography simulator based on Abbe's method is essential. {A graphical comparison of the imaging techniques of Hopkins and Abbe is provided in \Cref{fig:abbe-vs-hopkins}.}

\textbf{Approximation models}.
Although the Hopkins diffraction model is more computationally efficient than the Abbe's model, the intensity distribution is tedious to evaluate.
To reduce the computational complexity of the Hopkins diffraction model, a set of approximation models of partially coherent imaging systems are implemented,
including singular value decomposition (SVD) model, Fourier series expansion (FSE) model, average coherent approximation (ACA) model.

% \todo{FSE, ACA equation.}

\begin{figure}[tb!]
    \centering
    \includegraphics[width=.76\linewidth]{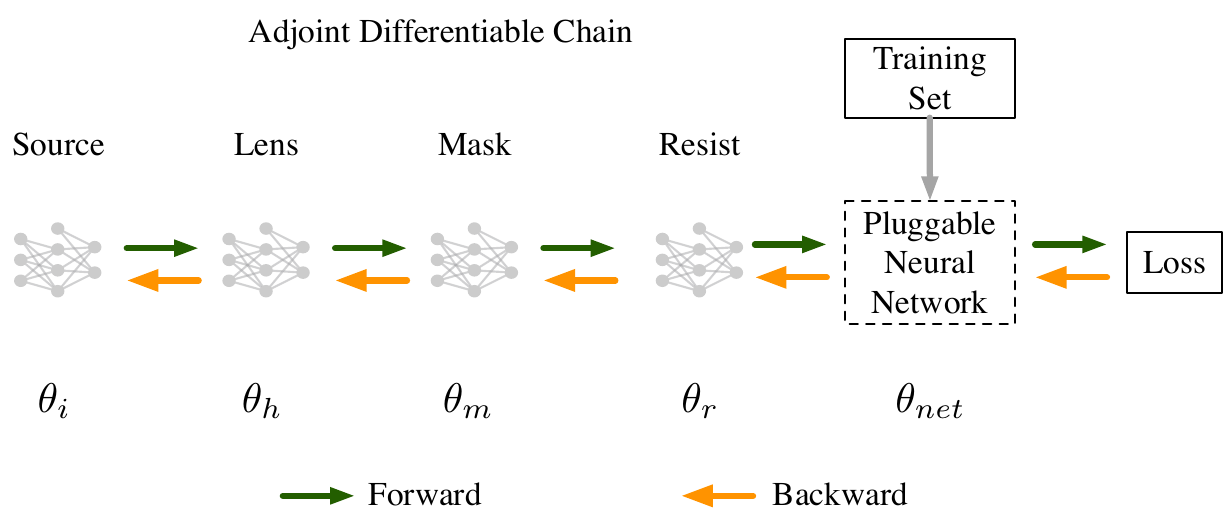}
    \caption{The visualization of the differentiable lithography chain.}
    \label{fig:dchain}
\end{figure}

The \textbf{Differentiable Computational Lithography Framework} endeavors to harness the potential of the differentiable programming paradigm and the parallel processing power of GPUs to enhance the precision of lithography modeling and simplify the optimization of RETs. A nuanced comprehension of the individual components constituting the encoding optical system is pivotal for crafting a more authentic model. The illumination setup, comprising light sources and potential optical constituents, emits photons that traverse either directly or by reflection, interacting with the intended object. Subsequently, the transmission and gathering optics, coupled with electronic sensors, amass information about the object carried by these wave patterns.
As illustrated in \Cref{fig:dchain}, each sub-process is characterized by its specific encoding function, denoted as \(f_{*}(\cdot, \mathbf{\theta})\), with \(\mathbf{\theta}\) symbolizing an array of parameters. The illumination component is depicted as a function \(f_i(\cdot, \mathbf{\theta}_i)\), influenced by factors like wavelength, coherence, and polarization. The projector's transfer functionality can be presented as a differentiable function \(f_h({\cdot, \mathbf{\theta}_h})\), governed by parameters like NA, reflection coefficient, focus, and intensity. Similarly, a two-dimensional tensor can depict the mask's function as \(f_m(\cdot, \mathbf{\theta}_m)\). Although typically this tensor appears as a binary matrix with primarily 0 or 1 values, for the sake of ensuring differentiability in computation, a matrix with continuous values, upon which an appropriate activation function is applied, is preferred. The resist model gets represented by \(f_r(\cdot, \mathbf{\theta}_r)\).
The comprehensive forward lithography model translates to a composite function as:
\begin{equation}
  f = \eta \circ f_i \circ f_{h} \circ f_m \circ f_r \Rightarrow Z = f(\cdot, \mathbf{\theta}) + \gamma,
\end{equation}
where \(\circ\) stands for the function composition mechanism, while \(Z\), \(\eta\), and \(\gamma\) denote final resist image, signal-dependent and signal-independent noises, respectively. The intricacy of an authentic forward model in imaging frameworks arises from each optical element's input combined with uncertainties due to system flaws and noise. Recognizing this complexity is crucial when fashioning image reconstruction methodologies or architecting imaging systems.
Given target image $Z_t$, once a comprehensive differentiable lithography mechanism is formulated, the model's parameters can be fine-tuned using gradient descent, culminating in a more refined lithography model as:
\begin{equation}
  (\theta_{i\_opt}, \theta_{h\_opt}) := \underset{\mathbf{\theta}_i, \mathbf{\theta}_h}{\operatorname{argmin}}|f(\cdot;\mathbf{\theta}_i, \mathbf{\theta}_h) - Z_t|^2.
\end{equation}

% \todo{Parameter tuning.}

% \todo{Maybe add algorithm flow into it.}

% \todo{Revise this paragraph}
\textbf{Adjoint back-propagation}, is a memory-efficient technique used in the context of differentiable computational lithography systems.
It is a method to compute gradients in a computational graph, which is particularly useful when dealing with large-scale optimization problems,
such as those involving complex optical systems and their associated imaging and optimization algorithms.
The standard back-propagation method in machine learning involves computing the gradients of a loss function with respect to the model parameters by traversing the computational graph from the output to the input.
However, this can be memory-intensive, especially when the graph is large or when dealing with millions of source or mask parameters in large scale VLSI.
Adjoint back-propagation, on the other hand, exploits the fact that the gradients can be computed in a more efficient way by separating the computation into multiple stages.
This separation allows for the computation of the gradients without the need to store intermediate states from the forward pass, which significantly reduces memory consumption.
In the context of the differentiable lithography engine introduced in this paper,
adjoint back-propagation is used to optimize the source shape, lens design, mask optimization and resist modeling algorithms jointly.
The method is applied to compute the gradients of the error metric with respect to the source, lens, mask parameters and the pluggable neural network modules.
This enables end-to-end optimization of both the optical system and the image processing algorithm, leading to improved image quality.
The adjoint back-propagation approach in differentiable lithography is particularly beneficial for computational lithography applications where the merit function is in the image space, such as mask optimization or source mask co-optimization.
By splitting the computation into multiple passes, differentiable lithography can handle large numbers of source illumination pixels and mask pixels,
which is crucial for practical end-to-end computational lithography designs.

\textbf{Objective functions}.
In optical lithography, light shines through a patterned mask onto a wafer.
The light alters the wafer's light-sensitive photoresist layer, which is then partially dissolved by solvents.
Historically, computational lithography marked its inception in the 1980s, primarily serving as a supplementary mechanism.
However, with the onset of the 180nm technology node, where the device's minimum line-width began to breach the exposure wavelength,
the optical proximity correction (OPC) emerged as an inextricable component, becoming an essential facet in photomask patterning.
As lithographic techniques embarked on a trajectory toward diminishing technology nodes, correction strategies witnessed iterative refinements.
The subsequent epoch saw an escalating reliance on resolution enhancement techniques (RETs)~\cite{DAC23_DiffPattern,SPIE23_GPU_MPL,ICCAD22_LayoutTransformer}, including but not limited to,
the meticulous optimization of illumination conditions, sub-resolution assist feature (SRAF), multiple patterning and OPC.
Pivotal breakthroughs, such as source mask optimization (SMO)~\cite{TCAD_DiffSMO} and the avant-garde inverse lithography technique (ILT) that emerged around 2010,
propelled computational lithography into an unprecedented echelon.
Notably, ILT, with its capacity to inferentially deduce mask patterns directly from wafer requisites, epitomizes the sophistication computational lithography has attained.
In the context of sub-32nm technology nodes, it is unambiguous that computational lithography has burgeoned to occupy a central position in lithographic R\&D endeavors.

The framework has introduced a diverse array of objective function toolkits specifically designed to cater to various RETs.
Central to this framework is the utilization of forward lithography models,
which are intricately structured around composition functions.
This approach not only enhances user interactivity but also fortifies the system's adaptability and flexibility.
To elucidate the framework's capabilities, we highlight two pivotal applications:
mask optimization, conventionally termed as OPC, and the SMO methodology.
Given the forward lithography modeling $f$, the objective function for OPC can be stated as
\begin{equation}
  {M}_{opt} = \underset{\mathbf{\theta}_m}{\operatorname{argmin}}|f(\cdot, \mathbf{\theta}_m) - Z_t|^2,
  \label{eq:opc_objective}
\end{equation}
where $M_{opt}$ is the optimized mask and $Z_t$ is the design target.
Since $f$ is differentiable, \Cref{eq:opc_objective} can be solved by gradient descent and various optimization methods.
Furthermore, in the context of addressing the source mask optimization (SMO) challenge,
it is plausible to employ a methodology analogous to OPC approaches.
However, a salient differentiation arises within the SMO paradigm:
there is an imperative to concurrently optimize both the mask parameters, denoted as $\theta_m$,
and the source parameters, represented as $\theta_i$.
Conventionally, the iterative process involves alternating the optimization endeavors between $\theta_m$ and $\theta_i$,
\begin{equation}
  (M_{opt}, J_{opt}) = \underset{\mathbf{\theta}_i, \mathbf{\theta}_m}{\operatorname{argmin}}|f(\cdot;\mathbf{\theta}_i, \mathbf{\theta}_m) - Z_t|^2,
\end{equation}
where $M_{opt}$, $J_{opt}$ are optimized mask and source, respectively.

\begin{table}[tb!]
  \centering
  \caption{Comparison among different lithography imaging models}
  % \resizebox{0.99\linewidth}{!}
  % {
    \renewcommand{\arraystretch}{1.1}
    \begin{tabular}{c|cc|cc|c|c|c}\toprule
      Model &\multicolumn{2}{c|}{Layer} &\multicolumn{2}{c|}{Imaging model} &Configurable? &GPU Acce. &Neural Network \\
      &Via &Metal &Abbe &Hopkins & & & \\\midrule
      LithoGAN~\cite{DFM-DAC2019-Ye} &\Checkmark &\XSolidBrush &- &- &\XSolidBrush &\Checkmark &GAN \\
      DAMO~\cite{ICCAD20_DAMO} &\Checkmark &\XSolidBrush &- &- &\XSolidBrush &\Checkmark &DCGAN \\
      DOINN~\cite{DAC22-DOINN-Yang} &\Checkmark &\Checkmark &- &\Checkmark &\XSolidBrush &\Checkmark &FNO + CNN \\
      ICCAD13~\cite{OPC-ICCAD2013-Banerjee} &\XSolidBrush &\Checkmark &- &\Checkmark &\XSolidBrush &\XSolidBrush &\XSolidBrush \\ \midrule
      \textbf{Ours} &\Checkmark &\Checkmark &\Checkmark &\Checkmark &\Checkmark &\Checkmark &Composable  \\
      \bottomrule
      \end{tabular}
  % }
  \label{tab:cmp_litho_methods}
\end{table}

\section{Results comparison}
\label{sec:result}  % \label{} allows reference to this section

%  We listed some preliminary results on both OPC and SMO tasks.
In the \Cref{tab:cmp_litho_methods}, we compare our method with the leading lithography imaging models found in academic research.
Using rigorous optical modeling, our approach is the first to support both the via layer and metal layer simultaneously.
We have incorporated both the Abbe and Hopkins models.
We also offer support for various approximation models.
Ours is the only method that allows users to adjust optical parameters. Lastly, we have integrated neural network technologies and provided support for GPU acceleration.

\begin{table}[htbp]
  \centering
  \caption{Result Comparison with State-of-the-Art.}
	\renewcommand{\arraystretch}{1.1}
  \begin{tabular}{l|cc|cc|cc|cc}\toprule
  \multirow{2}{*}{Dataset} &\multicolumn{2}{c|}{DAMO~\cite{ICCAD20_DAMO}} &\multicolumn{2}{c|}{TEMPO~\cite{TEMPO_ISPD}} &\multicolumn{2}{c|}{DOINN~\cite{DAC22-DOINN-Yang}} &\multicolumn{2}{c}{Ours} \\\cmidrule{2-9}
  &mPA &mIOU &mPA &mIOU &mPA &mIOU &mPA &mIOU \\\midrule
  Benchmark1~\cite{OPC-ICCAD2013-Banerjee} &95.2 &91.1 &94.6 &88.7 &99.19 &98.32 &99.45 &99.21 \\
  Benchmark2~\cite{ispd2019-benchmark} &98.97 &97.31 &98.24 &96.55 &98.79 &97.1 &99.15 &99.02 \\
  Benchmark3~\cite{ispd2019-benchmark} &99.11 &93.56 &99.06 &93.28 &99.21 &98.41 &99.59 &99.34 \\
  Benchmark4~\cite{OPC-ICCAD2013-Banerjee,ispd2019-benchmark} &99.01 &97.1 &98.63 &95.84 &98.71 &96.68 &99.61 &99.36 \\ \midrule
  Average &98.07 &94.77 &97.63 &93.59 &98.98 &97.63 &\textbf{99.45} &\textbf{99.23} \\
  Ratio &0.99 &0.96 &0.98 &0.94 &0.99 &0.98 &\textbf{1} &\textbf{1} \\
  \bottomrule
  \end{tabular}
  \label{tab:results}
\end{table}

% use more detailed
%We adopted four datasets from ICCAD~\cite{OPC-ICCAD2013-Banerjee} and ISPD~\cite{ispd2019-benchmark} contests and used Calibre to obtain the ground truth for lithography. When compared with the state-of-the-art (SOTA) lithography models,
%it can be observed that our model achieves the best results in both mIOU (Mean Intersection Over Union) and mPA (Mean Pixel Accuracy).

In this study, we utilized four benchmarks from the ICCAD~\cite{OPC-ICCAD2013-Banerjee} and ISPD~\cite{ispd2019-benchmark} contests, employing Calibre to derive the ground truth data essential for lithography simulation. These datasets, recognized for their comprehensive coverage and challenging nature, serve as a robust foundation for evaluating the performance of lithography predictive models. Our comparative analysis extends across several leading-edge models in the domain of advanced lithography simulation, including DAMO, TEMPO~\cite{TEMPO_ISPD}, and DOINN, alongside our proposed model.
The results, as detailed in the accompanying table, underscore the superior performance of our model across all benchmarks. Specifically, in terms of Mean Pixel Accuracy (mPA) and Mean Intersection Over Union (mIOU), two critical metrics for assessing the fidelity of lithographic simulations, our model consistently outperforms the competing approaches.
As shown in \Cref{tab:results}, on Benchmark1~\cite{OPC-ICCAD2013-Banerjee}, our model achieves an mPA of 99.45\% and an mIOU of 99.21\%,
compared to the next best model, DOINN, which records 99.19\% and 98.32\%, respectively. This trend of excellence is sustained across all datasets, culminating in an average mPA of 99.45\% and an mIOU of 99.23\% for our model, compared to the averages of the competing models which lag behind, highlighting the effectiveness and accuracy of our approach in simulating lithographic processes.
Moreover, the ratio of our model's performance against the benchmarks further emphasizes its optimization and capability in handling the intricacies of lithography simulation with unparalleled precision.
Our model sets a new standard for accuracy and reliability in the field, indicating a perfect alignment with the ground truth data obtained through Calibre.
These findings not only showcase the superior performance of our model over the state-of-the-art lithography models but also highlight its potential to significantly improve the accuracy and efficiency of lithography simulation processes, a crucial aspect for the advancement of semiconductor manufacturing technologies.

\begin{figure}[tb!]
  \centering
  \subfloat[]{\includegraphics[width=.42\linewidth]{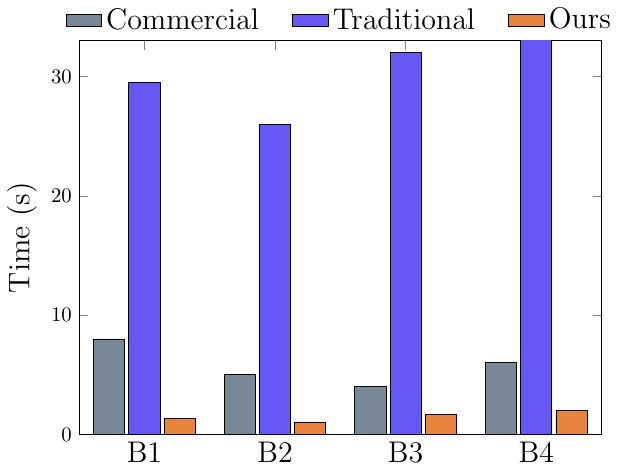} \label{fig:bar-graph}} \hspace{2em}
  \subfloat[]{\includegraphics[width=.42\linewidth]{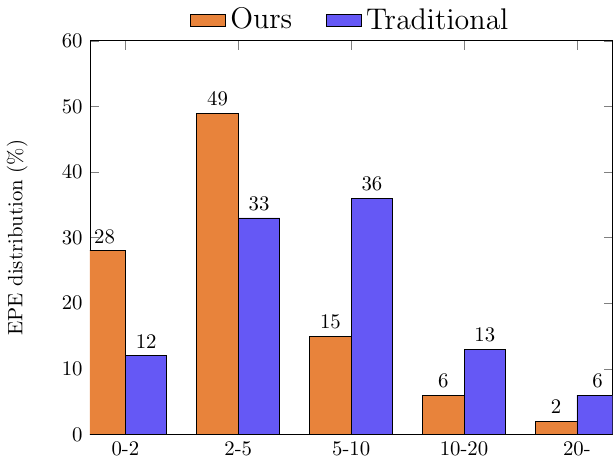} \label{fig:bar-epe}}
  \caption{
    (a) The result comparisons on runtime;
    (b) The result comparisons on EPE.
  }
  \label{fig:epe_rutime}
\end{figure}

% This bar graph presents a clear and comprehensive comparison of processing times across three different methods: Commercial, Traditional, and Ours, across four benchmarks labeled B1, B2, B3, and B4. The vertical axis quantifies the time in seconds required by each method to complete the task, providing a straightforward measure of efficiency.
In our comparative analysis of runtime performance and EPE distribution, as detailed in \Cref{fig:epe_rutime}, our methodology significantly outperforms traditional approaches in efficiency and precision.
The runtime evaluation, presented in \Cref{fig:bar-graph}, highlights our method's superior processing speeds across all benchmarks, with recorded times substantially lower than those of traditional methods, thereby underscoring its potential to markedly decrease processing durations.
This improvement in processing time is indicative of enhanced productivity and operational efficiency, particularly beneficial for applications demanding rapid processing capabilities.
Furthermore, the EPE distribution analysis, depicted in \Cref{fig:bar-epe}, illustrates our method's precision, especially within the critical 0-2 range where the majority of our results are concentrated, suggesting a high degree of accuracy.
This is in stark contrast to the traditional approach, which demonstrates a less favorable error distribution, with a notable proportion of errors falling within higher ranges, thereby indicating lower precision.
The contrast in EPE distribution between our method and traditional approaches emphasizes the significant reduction in average EPE achieved through our methodology.
By focusing the majority of errors within the most favorable 0-2 range, our approach not only affirms its accuracy but also its capacity to improve the reliability and performance of precision-dependent applications,
highlighting its superiority in both runtime efficiency and error precision.

\section{Conclusion}
\label{sec:conclu}  % \label{} allows reference to this section
In summary, this paper introduces a novel differentiable lithography imaging framework, showcasing its transformative potential in computational lithography.
Through the integration of differentiable programming principles, our framework offers enhanced precision and efficiency in semiconductor manufacturing.
Notably, it accurately models standard scalar imaging models like the Abbe and Hopkins models, demonstrating superior performance compared to existing approaches.
Additionally, its capacity to handle large-scale optimization problems with memory efficiency, facilitated by adjoint back-propagation, positions it as a robust tool for optical systems optimization.
The decision to open-source our models underscores our dedication to transparency and collaborative research efforts, with the aim of cultivating an enriched environment for lithography research.
While it is anticipated that this framework will contribute to advancements in computational lithography, aiding in the miniaturization of electronic devices and fostering the development of next-generation technologies,
its precise impact remains to be seen.
Nevertheless, the introduction of this differentiable lithography imaging framework marks a notable progression in computational lithography,
holding potential for further advancements in semiconductor manufacturing and supporting the pursuit of Moore's law.
The source code is available at \url{https://github.com/TorchOPC/TorchLitho}.

% \acknowledgments % equivalent to \section*{ACKNOWLEDGMENTS

% References
\bibliographystyle{spiebib} % makes bibtex use spiebib.bst
\bibliography{ref/sim.bib,ref/litho.bib,ref/DFM.bib,ref/OPC.bib,ref/SMO.bib} % bibliography data in report.bib

\begin{thebibliography}{10}

\bibitem{ma2011computational}
Ma, X. and Arce, G.~R.,  [{\em Computational
  lithography}{\nolinebreak\hspace{0.1em}]}, John Wiley \& Sons (2011).

\bibitem{DATE21_GPU_LevelSet}
Yu, Z., Chen, G., Ma, Y., and Yu, B., ``A gpu-enabled level set method for mask
  optimization,'' in [{\em IEEE/ACM Proceedings Design, Automation and Test in
  Eurpoe (DATE)}{\nolinebreak\hspace{0.1em}]},   1835--1838 (2021).

\bibitem{OPC-ICCAD2021-DevelSet}
Chen, G., Yu, Z., Liu, H., Ma, Y., and Yu, B., ``{DevelSet: Deep neural level
  set for instant mask optimization},'' in [{\em IEEE/ACM International
  Conference on Computer-Aided Design (ICCAD)}{\nolinebreak\hspace{0.1em}]},
  (2021).

\bibitem{DAC23_Nitho}
Chen, G., Pei, Z., Yang, H., Ma, Y., Yu, B., and Wong, M., ``Physics-informed
  optical kernel regression using complex-valued neural fields,'' in [{\em
  ACM/IEEE Design Automation Conference (DAC)}{\nolinebreak\hspace{0.1em}]},
  (2023).

\bibitem{ICCAD22_AdaOPC}
Zhao, W., Yao, X., Yu, Z., Chen, G., Ma, Y., Yu, B., and Wong, M. D.~F.,
  ``{AdaOPC}: A self-adaptive mask optimization framework for real design
  patterns,'' in [{\em Proceedings of the 41st IEEE/ACM International
  Conference on Computer-Aided Design}{\nolinebreak\hspace{0.1em}]},  (2022).

\bibitem{L2OILT_TCAD}
Zhu, B., Zheng, S., Yu, Z., Chen, G., Ma, Y., Yang, F., Yu, B., and Wong, M.,
  ``L2o-ilt: Learning to optimize inverse lithography techniques,'' {\em IEEE
  Transactions on Computer-Aided Design of Integrated Circuits and Systems}
  (09 2023).

\bibitem{liu2022adversarial}
Liu, M., Yang, H., Li, Z., Sastry, K., Mukhopadhyay, S., Dogru, S., Anandkumar,
  A., Pan, D.~Z., Khailany, B., and Ren, H., ``An adversarial active
  sampling-based data augmentation framework for manufacturable chip design,''
  {\em arXiv preprint arXiv:2210.15765}  (2022).

\bibitem{large_scale_phys_litho23}
Yang, H. and Ren, H., ``Enabling scalable ai computational lithography with
  physics-inspired models,'' in [{\em 2023 28th Asia and South Pacific Design
  Automation Conference (ASP-DAC)}{\nolinebreak\hspace{0.1em}]},   715--720
  (2023).

\bibitem{TCAD_DAMO}
Chen, G., Chen, W., Sun, Q., Ma, Y., Yang, H., and Yu, B., ``Damo: Deep agile
  mask optimization for full-chip scale,'' {\em IEEE Transactions on
  Computer-Aided Design of Integrated Circuits and Systems}~{\bf 41}(9),
  3118--3131 (2022).

\bibitem{Zhu22DNF}
Zhu, H., Liu, Z., Zhou, Y., Ma, Z., and Cao, X., ``Dnf: diffractive neural
  field for lensless microscopic imaging,'' {\em Opt. Express}~{\bf 30},
  18168--18178 (May 2022).

\bibitem{olah2015neural}
Olah, C., ``Neural networks, types, and functional programming,'' (2015).

\bibitem{tensorflow}
Abadi, M., Agarwal, A., Barham, P., Brevdo, E., Chen, Z., Citro, C., Corrado,
  G., Davis, A., Dean, J., Devin, M., Ghemawat, S., Goodfellow, I., Harp, A.,
  Irving, G., Isard, M., Jia, Y., Jozefowicz, R., Kaiser, L., Kudlur, M.,
  Levenberg, J., Mané, D., Monga, R., Moore, S., Murray, D., Olah, C.,
  Schuster, M., Shlens, J., Steiner, B., Sutskever, I., Talwar, K., Tucker, P.,
  Vanhoucke, V., Vasudevan, V., Viégas, F., Vinyals, O., Warden, P.,
  Wattenberg, M., Wicke, M., Yu, Y., and Zheng, X., ``Tensorflow: Large-scale
  machine learning on heterogeneous distributed systems,'' (2015).

\bibitem{pytorch}
Paszke, A., Gross, S., Massa, F., Lerer, A., Bradbury, J., Chanan, G., Killeen,
  T., Lin, Z., Gimelshein, N., Antiga, L., Desmaison, A., K\"{o}pf, A., Yang,
  E., DeVito, Z., Raison, M., Tejani, A., Chilamkurthy, S., Steiner, B., Fang,
  L., Bai, J., and Chintala, S.,  [{\em PyTorch: An Imperative Style,
  High-Performance Deep Learning Library}{\nolinebreak\hspace{0.1em}]}, Curran
  Associates Inc., Red Hook, NY, USA (2019).

\bibitem{jax2018github}
Bradbury, J., Frostig, R., Hawkins, P., Johnson, M.~J., Leary, C., Maclaurin,
  D., Necula, G., Paszke, A., Vander{P}las, J., Wanderman-{M}ilne, S., and
  Zhang, Q., ``{JAX}: composable transformations of {P}ython+{N}um{P}y
  programs,'' (2018).

\bibitem{chen2023differentiable}
Chen, N., Cao, L., Poon, T.-C., Lee, B., and Lam, E.~Y., ``Differentiable
  imaging: A new tool for computational optical imaging,'' {\em Advanced
  Physics Research}~{\bf 2}(6),  2200118 (2023).

\bibitem{dreamplace}
Lin, Y., Dhar, S., Li, W., Ren, H., Khailany, B., and Pan, D.~Z.,
  ``{DREAMPIace}: Deep learning toolkit-enabled gpu acceleration for modern
  vlsi placement,'' in [{\em 2019 56th ACM/IEEE Design Automation Conference
  (DAC)}{\nolinebreak\hspace{0.1em}]},   1--6 (2019).

\bibitem{dreamplace3}
Gu, J., Jiang, Z., Lin, Y., and Pan, D.~Z., ``{DREAMPlace} 3.0:
  Multi-electrostatics based robust vlsi placement with region constraints,''
  in [{\em 2020 IEEE/ACM International Conference On Computer Aided Design
  (ICCAD)}{\nolinebreak\hspace{0.1em}]},   1--9 (2020).

\bibitem{ICCAD23:AlphaSyn}
Pei, Z., Liu, F., He, Z., Chen, G., Zheng, H., Zhu, K., and Yu, B., ``Alphasyn:
  Logic synthesis optimization with efficient monte carlo tree search,'' in
  [{\em IEEE/ACM International Conference on Computer-Aided Design
  (ICCAD)}{\nolinebreak\hspace{0.1em}]},  (11 2023).

\bibitem{ECCV22_hilbert}
Chen, W., Zhu, X., Chen, G., and Yu, B., ``Efficient point cloud analysis using
  hilbert curve,'' in [{\em European Conference on Computer Vision
  (ECCV)}{\nolinebreak\hspace{0.1em}]},  (2022).

\bibitem{ISPD21_Pointcloud}
Li, W., Chen, G., Yang, H., Chen, R., and Yu, B., ``Learning point clouds in
  eda,'' in [{\em Proceedings of the 2021 International Symposium on Physical
  Design}{\nolinebreak\hspace{0.1em}]},  (2021).

\bibitem{ICCAD20_DAMO}
Chen, G., Chen, W., Ma, Y., Yang, H., and Yu, B., ``{DAMO}: Deep agile mask
  optimization for full chip scale,'' in [{\em Proceedings of the 39th
  International Conference on Computer-Aided
  Design}{\nolinebreak\hspace{0.1em}]},  (2020).

\bibitem{fuhner2014artificial}
F{\"u}hner, T., {\em Artificial evolution for the optimization of lithographic
  process conditions}, PhD thesis, Friedrich-Alexander-Universit{\"a}t
  Erlangen-N{\"u}rnberg (FAU) (2014).

\bibitem{OPC-DAC2014-Gao}
Gao, J.-R., Xu, X., Yu, B., and Pan, D.~Z., ``{MOSAIC}: Mask optimizing
  solution with process window aware inverse correction,'' in [{\em ACM/IEEE
  Design Automation Conference (DAC)}{\nolinebreak\hspace{0.1em}]},
  52:1--52:6 (2014).

\bibitem{OPC-DAC2018-Yang}
Yang, H., Li, S., Ma, Y., Yu, B., and Young, E.~F., ``{GAN-OPC}: Mask
  optimization with lithography-guided generative adversarial nets,'' in [{\em
  ACM/IEEE Design Automation Conference (DAC)}{\nolinebreak\hspace{0.1em}]},
  131:1--131:6 (2018).

\bibitem{a2ilt2022wang}
Wang, Q., Jiang, B., Wong, M. D.~F., and Young, E. F.~Y., ``{A2-ILT}: Gpu
  accelerated ilt with spatial attention mechanism,'' in [{\em ACM/IEEE Design
  Automation Conference (DAC)}{\nolinebreak\hspace{0.1em}]},  (2022).

\bibitem{multiilt2023sun}
Sun, S., Yang, F., Yu, B., Shang, L., and Zeng, X., ``Efficient {ILT} via
  multi-level lithography simulation,'' in [{\em ACM/IEEE Design Automation
  Conference (DAC)}{\nolinebreak\hspace{0.1em}]},  (2023).

\bibitem{DAC23_DiffPattern}
Wang, Z., Shen, Y., Zhao, W., Bai, Y., Chen, G., Farnia, F., and Yu, B.,
  ``{DiffPattern}: Layout pattern generation via discrete diffusion,'' in [{\em
  ACM/IEEE Design Automation Conference (DAC)}{\nolinebreak\hspace{0.1em}]},
  (2023).

\bibitem{SPIE23_GPU_MPL}
Chen, G., Yang, H., and Yu, B., ``{GPU-accelerated matrix cover algorithm for
  multiple patterning layout decomposition},'' in [{\em DTCO and Computational
  Patterning II}{\nolinebreak\hspace{0.1em}]},  International Society for
  Optics and Photonics, SPIE (2023).

\bibitem{ICCAD22_LayoutTransformer}
Wen, L., Zhu, Y., Ye, L., Chen, G., Yu, B., Liu, J., and Xu, C.,
  ``{LayouTransformer}: Generating layout patterns with transformer via
  sequential pattern modeling,'' in [{\em Proceedings of the 41st IEEE/ACM
  International Conference on Computer-Aided
  Design}{\nolinebreak\hspace{0.1em}]},  (2022).

\bibitem{TCAD_DiffSMO}
Chen, G., Wang, Z., Yu, B., Pan, D.~Z., and Wong, M.~D., ``Ultra-fast source
  mask optimization via conditional discrete diffusion,'' {\em IEEE
  Transactions on Computer-Aided Design of Integrated Circuits and Systems} ,
  1--1 (2024).

\bibitem{DFM-DAC2019-Ye}
Ye, W., Alawieh, M.~B., Lin, Y., and Pan, D.~Z., ``{LithoGAN}: End-to-end
  lithography modeling with generative adversarial networks,'' in [{\em
  ACM/IEEE Design Automation Conference (DAC)}{\nolinebreak\hspace{0.1em}]},
  107:1--107:6 (2019).

\bibitem{DAC22-DOINN-Yang}
Yang, H., Li, Z., Sastry, K., Mukhopadhyay, S., Kilgard, M., Anandkumar, A.,
  Khailany, B., Singh, V., and Ren, H., ``Generic lithography modeling with
  dual-band optics-inspired neural networks,'' in [{\em ACM/IEEE Design
  Automation Conference (DAC)}{\nolinebreak\hspace{0.1em}]},   973–978
  (2022).

\bibitem{OPC-ICCAD2013-Banerjee}
Banerjee, S., Li, Z., and Nassif, S.~R., ``{ICCAD-2013 CAD} contest in mask
  optimization and benchmark suite,'' in [{\em IEEE/ACM International
  Conference on Computer-Aided Design (ICCAD)}{\nolinebreak\hspace{0.1em}]},
  271--274 (2013).

\bibitem{TEMPO_ISPD}
Ye, W., Alawieh, M.~B., Watanabe, Y., Nojima, S., Lin, Y., and Pan, D.~Z.,
  ``Tempo: Fast mask topography effect modeling with deep learning,'' in [{\em
  Proceedings of the 2020 International Symposium on Physical
  Design}{\nolinebreak\hspace{0.1em}]},  {\em ISPD '20},  127–134,
  Association for Computing Machinery, New York, NY, USA (2020).

\bibitem{ispd2019-benchmark}
Liu, W.-H., Mantik, S., Chow, W.-K., Ding, Y., Farshidi, A., and Posser, G.,
  ``Ispd 2019 initial detailed routing contest and benchmark with advanced
  routing rules,'' in [{\em Proceedings of the 2019 International Symposium on
  Physical Design}{\nolinebreak\hspace{0.1em}]},   147–151 (2019).

\end{thebibliography}

\end{document}